\documentclass{elsart}

\usepackage{epsfig}

\usepackage{amssymb}

\begin{document}
\begin{frontmatter}




\title{On the interpretation of the equilibrium magnetization in 
the mixed state of high-$T_c$ superconductors.}

\author[label1,label2]{I. L. Landau}
\author[label1]{H. R. Ott}
 \address[label1]{Laboratorium f\"{u}r Festk\"{o}rperphysik, 
 ETH-H\"{o}nggerberg, CH-8093 Z\"{u}rich, Switzerland}
 \address[label2]{Institute for Physical Problems, 117334 Moscow, 
Russia}



\begin{abstract}
	
We apply a recently developed scaling procedure to the analysis of 
equilibrium magnetization $M(H)$ data that were obtained for 
Tl$_2$Ba$_2$CaCu$_2$O$_{8+x}$ and Bi$_2$Sr$_2$CaCu$_2$O$_8$ 
single crystals and were reported in the literature. The temperature 
dependencies of the upper critical field and the magnetic field 
penetration depth resulting from our analysis are distinctly 
different from those obtained in the original publications. We argue 
that theoretical models, which are usually employed for analyses of 
the equilibrium magnetization in the mixed state of type-II 
superconductors are not adequate for a quantitative description of 
high-$T_{c}$ superconductors. In addition, we demonstrate that the 
scaled equilibrium magnetization $M(H)$ curve for a Tl-2212 sample 
reveals a pronounced kink, suggesting a phase transition in the 
mixed state. 

\end{abstract}

\end{frontmatter}


Measurements of the equilibrium magnetization in the mixed state of 
type-II superconductors are often used for studying conventional 
and unconventional superconductivity. This is particularly true for 
high-$T_c$ (HTSC) superconductors, because of the extremely wide 
range of magnetic fields in which their magnetization is reversible. 
The physically meaningful information is not straightforwardly 
accessible, however, and, in order to estimate critical magnetic 
fields or characteristic lengths from magnetization measurements, 
theoretical models have to be invoked. Below we consider several 
theoretical approaches which are usually employed for the 
interpretation of corresponding experimental results, and show that 
the resulting temperature dependencies of the upper critical field 
$H_{c2}$ and the magnetic field penetration depth $\lambda$ still 
leave space for improvement.

The Hao-Clem model \cite{2,3} is most widely used for evaluating 
essential superconducting parameters, such as the upper critical 
field $H_{c2}$ and the Ginzburg-Landau parameter $\kappa$ of HTSC's 
from magnetization data. Because this model takes into account the 
spatial variation of the order parameter it is, no doubt, a better 
approximation to the Abrikosov theory of the mixed state 
\cite{abrikos} than previously used approaches. Nevertheless, 
the $\kappa (T)$ curves obtained by employing the Hao-Clem model 
practically always exhibit a rather strong and unphysical increase 
of $\kappa$ with increasing temperature 
\cite{4,hc1,hc2,hc3,hc4,hc5,hc6,hc7,hc8,hc9,hc10,hc11,hc12}. 
\footnote{In the Ginzburg-Landau theory, $\kappa$ is temperature 
independent, and a slight reduction of $\kappa$ with increasing 
temperature is predicted by microscopic theories \cite{th1,th2}.} 
An instructive example of this behavior is provided by Figs. 3(a) 
and 3(b) in Ref. \cite{4} where equilibrium magnetization data of a 
Tl$_2$Ba$_2$CaCu$_2$O$_{8+x}$ (Tl-2212) single crystal were 
presented and analyzed. The quoted figures reveal a rather strong 
increase of $\kappa$ with increasing temperature and, as a 
consequence, a very unusual temperature dependence of the upper 
critical field $H_{c2}$, which is shown in the bottom inset of 
Fig. 1. As may be seen, $H_{c2}(T)$ resulting for this particular 
sample is  temperature independent below $T \approx 73$ K and 
exhibits an unphysical divergence at higher temperatures. 
A very similar behavior of the $H_{c2}(T)$ curve, resulting from 
the same type of analysis, was also reported for 
Bi$_2$Sr$_2$CaCu$_2$O$_8$ (Bi-2212) samples \cite{hc2}. More recent 
theoretical work \cite{blk,kogan2,kogan} was intended to avoid 
these inadequacies by taking into account some specific 
corrections to the sample magnetization that are not accounted for 
in the Abrikosov theory. However, as we argue below, the situation 
concerning the temperature dependencies of $H_{c2}$ and $\lambda$ 
may be improved even further by employing a recently established 
scaling procedure \cite{5}. 
\begin{figure}[t!]
 \begin{center}
  \epsfxsize=0.75\columnwidth \epsfbox {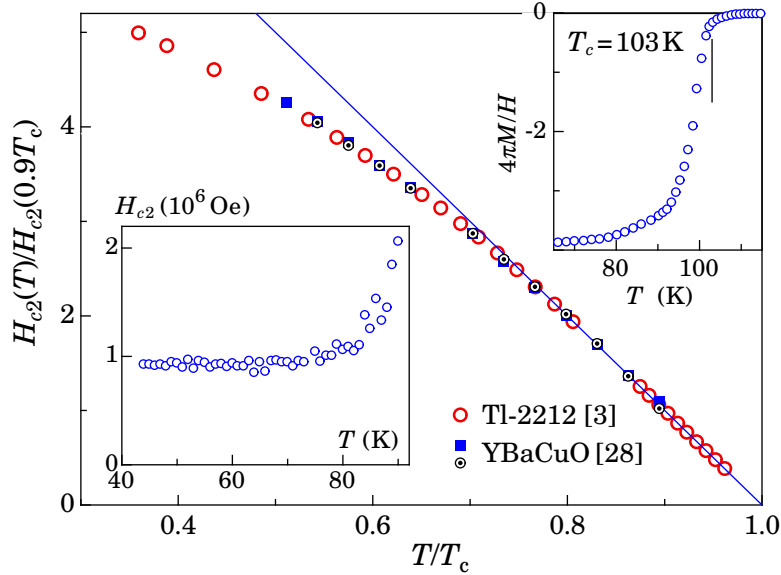}
  \caption{The temperature dependence of the normalized upper 
           critical field resulting from our scaling procedure for 
           Tl-2212 sample studied in Ref. \cite{4}. The solid line 
           is the best linear fit to the data points for $T \ge 90$ K. 
           The data points for two underdoped YBCO samples, 
           investigated in Ref. \cite{10}, were taken from Ref. 
           \cite{ceramics}. The upper inset shows the $M(T)$ curve 
           measured at $H = 10$ Oe. The short vertical line indicates 
           the position of $T_{c}$ as evaluated by the linear 
           extrapolation of $h_{c2}(T)$ to $h_{c2} = 0$. The bottom 
           inset shows $H_{c2}(T)$ as 
           obtained in Ref. \cite{4} by employing the Hao-Clem model.}
 \end{center}
\end{figure}

Our analysis of original data that were presented and discussed in 
Refs. \cite{4,hc2,kogan2} is based on a scaling procedure developed 
in Ref. \cite{5}. It was shown that if the Ginzburg-Landau parameter 
$\kappa$ is temperature independent, the equilibrium magnetizations 
in mixed state of type-II superconductors at two different 
temperatures are related by 
\begin{equation}
M(H/h_{c2},T_0)=M(H,T)/h_{c2}+c_0(T)H
\end{equation}
with
\begin{equation}
c_0(T)=\chi_{eff}^{(n)}(T)-\chi_{eff}^{(n)}(T_0).
\end{equation}
In Eqs. (1) and (2) $h_{c2}(T) = H_{c2}(T)/H_{c2}(T_{0})$ represents 
the normalized upper critical field and $\chi_{eff}^{(n)}(T)$ is the 
effective magnetic susceptibility of the superconductor in the normal 
state. $T_0$ is an arbitrary chosen temperature within the covered 
range of temperatures. The first term on the right-hand side of Eq. 
(1) is universal for any type-II superconductor, while the second is 
introduced o account for the temperature dependent paramagnetism of 
HTSC's in the normal state. In practice, this second term often also 
includes a non-negligible contribution arising from the sample holder. 
In the following we use $M_{eff}(H)=M(H,T_0)$ to denote the 
magnetization calculated from measurements at $T \ne T_{0}$ using Eq. 
(1). The adjustable parameters $h_{c2}(T)$ and $c_{0}(T)$ may be 
established from the condition that the $M_{eff}$ curves, calculated 
from measured $M(H)$ data in the reversible regime at different 
temperatures, collapse onto a single $M_{eff}(H)$ curve which 
represents the equilibrium magnetization at $T = T_{0}$ (see Ref. 
\cite{5} for details). 

\begin{figure}[t]
 \begin{center}
  \epsfxsize=0.75\columnwidth \epsfbox {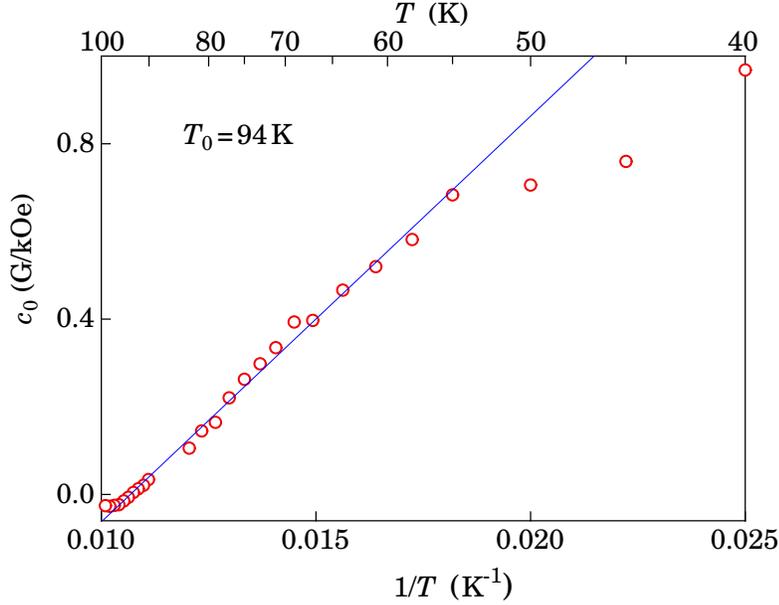}
  \caption{The scaling parameter $c_0$ as a function of $1/T$.} 
 \end{center}
\end{figure}
There are two adjustable parameters in our scaling procedure whereby 
the upper critical field $H_{c2}$ represents the natural 
normalization parameter for all magnetic characteristics of the 
mixed state and, as stated, $c_0(T)$ is essential to account for any 
temperature dependence of the paramagnetic susceptibility of HTSC's 
in the normal 
state. An important advantage of our scaling approach is that no 
particular field dependence of the magnetization has to be assumed 
{\it a priori} and therefore, this procedure may be used for any 
type-II superconductor, independent of the pairing type, the absolute 
value of $\kappa$, the anisotropy of superconducting parameters, or 
the sample geometry. However, because $M(H)$ is not postulated, the 
scaling procedure may only provide the relative temperature variation 
of $H_{c2}$ given by the scaling parameter $h_{c2}(T)$. The success 
of the scaling procedure described by  Eq. (1) in data analyses of a 
number of different HTSC materials was demonstrated in previous work 
\cite{5,jltp,ceramics,T,prb}. In addition to the temperature 
dependence of the normalized upper critical field, also the 
superconducting critical temperature can be evaluated by 
extrapolation of the $h_{c2}(T)$ curve to $h_{c2}=0$. The $T_c$ 
values evaluated in this way are always consistent with 
low field $M(T)$ curves \cite{5,ceramics}.  

The normalized upper critical field $h_{c2}(T)$, obtained via the 
scaling of $M(H)$ data for a Tl-2212 single crystal presented in
\cite{4}, is shown in Fig. 1. As may be seen, $h_{c2}(T)$ for this 
sample varies linearly with $T$ above $0.8T_c \approx 82$ K. This 
linearity allows for a quite accurate evaluation of the critical 
temperature $T_{c}$ by extrapolating the $h_{c2}(T)$ curve to 
$h_{c2} = 0$. The inset of Fig. 1 demonstrates that the value of 
$T_{c} = 103.0$ K resulting from this extrapolation, is well in 
agreement with the temperature dependence of the low-field 
magnetization of the same sample. As may be seen in Fig. 1, 
$h_{c2}(T/T_{c})$ for this Tl-2212 sample is practically identical 
with the $h_{c2}(T/T_{c})$ curves, presented in Ref. \cite{ceramics}, 
for two underdoped YBa$_2$Cu$_3$O$_{7-x}$ (YBCO) samples thus 
supporting our previous suggestion \cite{ceramics} concerning the 
universality of the temperature dependence of $H_{c2}(T)$ for HTSC's.

The temperature dependence of the scaling parameter $c_0$, which is 
shown in Fig. 2, demonstrates that the paramagnetic contribution to 
the sample magnetization obeys a Curie-type law in a rather extended 
temperature range with some deviations at temperatures below 55 K, 
as well as at temperatures very close to $T_c$.

The dependence of $M_{eff}$ on $H/h_{c2}(T)$ that results from our 
scaling procedure is shown in Fig. 3. Because of the high quality 
of the experimental data presented in Ref. \cite{4} and the 
extended covered range of magnetic fields, the scaling is nearly 
perfect. The $M_{eff}(H)$ data points, calculated from the 
measurements at different temperatures, combine to a single curve 
with virtually no scatter. The remarkable feature of this curve is 
a pronounced kink at $H/h_{c2} \approx 20$ kOe. This kink clearly 
indicates a significant change in the properties of the mixed state. 
Unfortunately, the measurements in Ref. \cite{4} are limited to 
magnetic fields $H \le 20$ kOe. Only a limited number of data 
points, measured at $T \ge 96$ K and $H = 20$ kOe combine to the 
$M_{eff}(H)$ curve above the kink. This is why, on the basis of the 
available data, no definite conclusions concerning this observation 
can be made. Only additional measurements in higher fields can 
clarify the situation. If this feature is confirmed by a more 
detailed study, the kink would  definitely reflect some kind of 
phase transition.

\begin{figure}[t]
 \begin{center}
  \epsfxsize=0.75\columnwidth \epsfbox {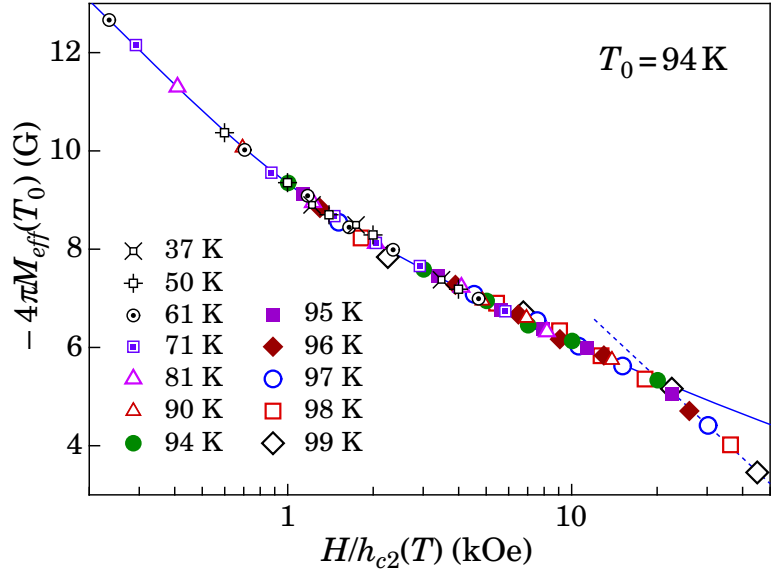}
  \caption{The scaled magnetization $M_{eff}$ calculated for $T_{0} 
           = 94$ K. The solid and the broken lines are meant to emphasize 
           the change of the slope.} 
 \end{center}
\end{figure}
If the Ginzburg-Landau parameter $\kappa$ is temperature 
independent, as it is assumed in our approach, the magnetic field 
penetration depth $\lambda (T)$ is inversely proportional to the 
square root of the upper critical magnetic field, i.e., 
\begin{equation}
\lambda (T)/\lambda (T_0) = \sqrt{H_{c2}(T_0)/H_{c2}(T)}.
\end{equation}
The temperature dependence of the normalized penetration depth 
calculated in this way is shown in Fig. 4. As may be seen in the 
inset of Fig. 4, for $T \geq 0.75T_c$ the resulting temperature 
dependence of $\lambda$ is well in agreement with the prediction of 
the Ginzburg-Landau theory. At lower temperatures, Eq. (3) is not 
applicable for the evaluation of $\lambda (T)$ and therefore it is 
not surprising that $\lambda (T)$ deviates from the Ginzburg-Landau 
type behavior. The plots in Fig. 4 also demonstrate that our 
$\lambda (T)$ curve is quite different from those calculated in 
Ref. \cite{4} by employing either a modified London model (nonlocal 
theory) \cite{kogan} or the Bulaevskii-Ledvji-Kogan approach (BLK 
theory) \cite{blk}. These differences are particularly pronounced 
in the temperature range where our $\lambda (T)$ curve matches the 
Ginzburg-Landau theory. 

\begin{figure}[t]
 \begin{center}
  \epsfxsize=0.75\columnwidth \epsfbox {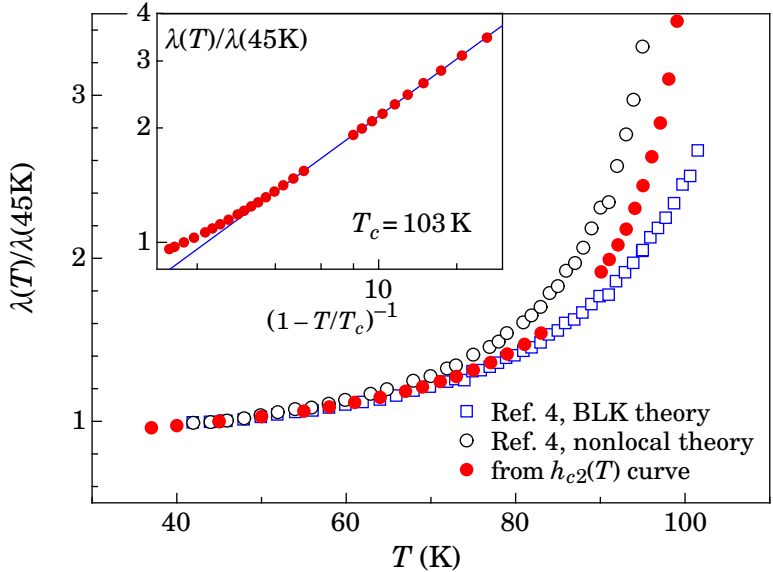}
  \caption{The normalized temperature dependence of $\lambda$ 
           calculated from the $h_{c2}(T)$ curve. The two $\lambda 
           (T)$ curves presented in Ref. \cite{4} are shown for 
           comparison. The inset shows the normalized $\lambda$ vs 
           $1/(1-T/T_c)$ on Log-scales; the straight line  
           corresponds to expectations of the Ginzburg-Landau theory.} 
 \end{center}
\end{figure}
Next we consider another example to support our arguments. We apply 
our analysis procedure on experimental $M(H)$ data for a single 
crystal of Bi$_2$Sr$_2$CaCu$_2$O$_8$, published in Ref. \cite{hc2}. 
These results were previously analyzed in Ref. \cite{kogan2} and we 
compare our results with those presented in \cite{kogan2}. In Fig. 5 
we show a comparison of the temperature dependence of the normalized 
upper critical field obtained from the same experimental data by 
employing three different approaches.

\begin{figure}[t]
 \begin{center}
  \epsfxsize=0.75\columnwidth \epsfbox {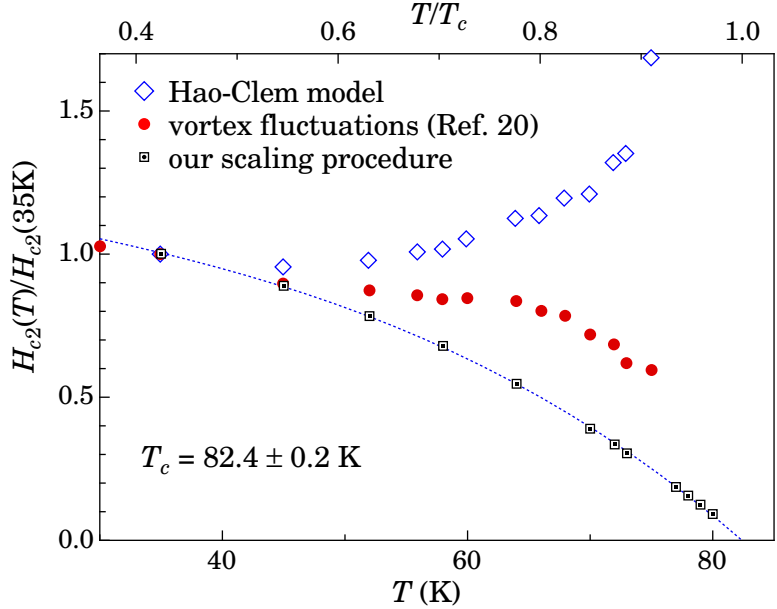}
  \caption{$H_{c2}(T)/H_{c2}$(35K) as a function of temperature. 
           These results are obtained by different analyses of the 
           same magnetization curves for a Bi$_2$Sr$_2$CaCu$_2$O$_8$ 
           sample which was experimentally investigated in Ref. 
           \cite{hc2}.} 
 \end{center}
\end{figure}
The plots in Figs. 1, 4 and 5, demonstrate that employing our 
scaling procedure for the analysis of the equilibrium magnetization 
data results in conventional temperature dependencies of $H_{c2}$ 
and $\lambda$ which are rather different from those obtained by 
invoking other, previously considered approaches 
\cite{2,3,blk,kogan2,kogan}. The models discussed in 
\cite{kogan2,kogan} were specially invented in order to explain the 
failure of the Hao-Clem model in handling magnetization data for 
layered HTSC compounds, as exemplified by the bottom inset of Fig. 1, 
and to improve the interpretation of experimental results. The 
enhancement of $H_{c2}$ at high temperatures was explained in Ref. 
\cite{kogan2} by the influence of thermal fluctuations on the sample 
magnetization, which are not accounted for in the Hao-Clem model. 
The temperature independence of $H_{c2}$ at lower temperatures was 
interpreted in Ref. \cite{kogan} in terms of non-local effects. It 
is argued in \cite{kogan} that due to effects of non-locality, which 
are expected to be important at temperatures well below $T_c$, the 
upper critical field $H_{c2}$ in the expression for the sample 
magnetization should be replaced by another field $H_0$ which 
depends on temperature much weaker than $H_{c2}$. The results of 
our analysis give no support to any of these two assumptions. As 
was demonstrated above, the sample magnetization can be scaled by 
simply invoking $H_{c2}(T)$ in a wide temperature range without 
significant corrections from thermal fluctuations or other sources.

Because the influence of thermal fluctuations on the magnetization 
of HTSC's in the mixed state is often overestimated in the 
literature, we discuss this point in more detail. As may clearly 
be seen in Fig. 3, if the experimental magnetizations are corrected 
for the temperature dependence of $\chi_{eff}^{(n)}$, the 
$M_{eff}(H)$ data points, corresponding to different temperatures, 
merge onto exactly the same curve. The obvious conclusion is that 
fluctuation effects do not produce a considerable contribution to 
the sample magnetization in this rather wide temperatures range. It 
should be noted, however that, as may be seen in Fig. 2, the 
$c_0(T)$ data points for the three highest temperatures ($T \ge 97$ 
K) deviate from the straight line corresponding to the Curie law. 
It is possible that these deviations are indeed due to thermal 
fluctuations.\footnote{As was pointed out in Ref. \cite{5}, at 
temperatures close to $T_c$, the term $c_0(T)H$ may also account 
for fluctuation effects. However, because the fluctuation-induced 
contribution to the sample magnetization is not linear in $H$, it 
may only approximately be accounted for.}  At still higher 
temperatures $T \ge 100$ K $\approx 0.97T_c$, the magnetization 
data presented in Ref. \cite{4} cannot satisfactorily be scaled 
using Eq. (1). We consider this failure of the scaling procedure as 
evidence for the increasing role of fluctuations with increasing 
temperature. We also note that in Y-123 compounds the impact of 
fluctuations effects is even weaker and in some cases our scaling 
procedure could successfully be employed up to temperatures as 
high as $0.99T_c$ \cite{5}. 

At present, it is difficult to identify the exact reason for the 
failure of the Hao-Clem model, an approximation to the 
Ginzburg-Landau theory of the mixed state, in treating experimental 
magnetization data. A few possibilities are mentioned below. 

1. According to Ref. \cite{pogosov}, the Hao-Clem model suffers from 
general inaccuracies in corresponding calculations. It is also 
quite possible that the contribution to the sample magnetization due 
to the normal-state paramagnetism of HTSC's is not accounted for with 
sufficient accuracy in the calculations.

2. Not all the assumptions of the model are satisfied in experiments. 
In particular, the assumed conventional $s-$pairing \cite{2,3}, also 
the basis for numerical calculations of Brandt \cite{brandt,brandt2}, 
is not compatible with the now accepted $d-$pairing in HTSC's.

3. More general reasons for the disagreement between the calculated 
and the experimental magnetization curves cannot be excluded. The 
argument is based on results of recent magnetization measurements on 
single crystals of NbSe$_2$ \cite{nbse}, a conventional superconductor 
without any expectation of unconventional pairing. Because the upper 
critical field for this superconductor is not very high, the 
paramagnetic contribution to the sample magnetization may easily be 
established and accounted for. For magnetic fields $H > 0.6H_{c2}$ 
(below $0.6H_{c2}$ the $M(H)$ curves are irreversible) the 
experimental magnetization $M(H)$ curves are linear if plotted versus 
$\ln H$. Very accurate numerical calculations of Brandt 
\cite{brandt,brandt2} for this magnetic field range result in $M(H)$ 
curves that vary linearly with $H$, however. 

In conclusion. by invoking published results of magnetization 
measurements on Tl-2212 \cite{4} and Bi-2212 \cite{hc2} samples we 
demonstrate that our analysis of equilibrium magnetization data 
results in temperature variations of $H_{c2}$ and $\lambda$ which 
are rather different from those obtained by employing theoretical 
approaches that are traditionally used for the interpretation of 
magnetization measurements in HTSC's \cite{2,3,blk,kogan2,kogan}. 
In view of the successful scaling of the magnetization data which 
is based on a minimum of {\it a priori} assumptions and whose 
validity was convincingly demonstrated in Refs. 
\cite{5,ceramics,T,nbse,thomp}, we are confident that 
the resulting temperature dependencies of the normalized upper 
critical field presented in Figs. 1 and 5 correctly describe the 
$H_{c2}(T)$ curves for these HTSC compounds.  We also argue that 
thermal fluctuations have a much weaker impact on the mixed-state 
magnetization of HTSC's than is usually believed.

\end{document}